\documentclass[a4paper,12pt]{article}

\begin{document}

\begin{flushright}
\begin{minipage}{3cm}
\begin{flushleft}
DPNU-03-11
\end{flushleft}
\end{minipage}
\end{flushright}

\begin{center}
 \large{\bf Violation of Vector Dominance
  in the Vector Manifestation}\footnote{
    Talk given at Fifth Symposium on 
    "Science of Hadrons under Extreme Conditions", 
    March 18-20, 2003, JAERI, Ibaraki, Japan.
    This talk is based on the work in Ref.~\cite{HS:VD}.}
\end{center}

\begin{center}
 Chihiro Sasaki \\
 \it{Department of Physics, Nagoya University, Nagoya, 464-8602, Japan}
\end{center}

\begin{abstract}
The vector manifestation (VM) is a new pattern for realizing
the chiral symmetry in QCD.
In the VM, the massless vector meson becomes the chiral partner 
of pion at the critical point, in contrast with the restoration 
based on the linear sigma model.
Including the intrinsic temperature
dependences of the parameters of the HLS Lagrangian determined from
the underlying QCD through the Wilsonian matching together with the
hadronic thermal corrections,
we present a new prediction of the VM on the
direct photon-$\pi$-$\pi$ coupling which measures the validity of the
vector dominance (VD) of the electromagnetic form factor of the pion.
We find that the VD is largely violated at the critical temperature,
which indicates that the assumption of the VD made in several analysis
on the dilepton spectra in hot matter may need to be weakened for
consistently including the effect of the dropping mass of the vector
meson.
\end{abstract}

\section{Introduction}

Spontaneous chiral symmetry breaking is one of the important  
features in low-energy QCD.
This symmetry breaking is expected to be restored in hot and/or
dense QCD and properties of hadrons will be changed near 
the critical temperature of the chiral symmetry restoration
~\cite{HatsudaKunihiro,Pisarski:95,%
Brown-Rho:96,Brown:2001nh,HatsudaShiomiKuwabara,%
Rapp-Wambach:00,Wilczek}.
The CERN Super Proton Synchrotron (SPS) observed
an enhancement of dielectron ($e^+e^-$) mass spectra
below the $\rho / \omega$ resonance.
This can be explained by the dropping mass of the $\rho$ meson
(see, e.g., Refs.~\cite{Li:1995qm, Brown-Rho:96, Rapp-Wambach:00})
following the Brown-Rho scaling proposed in Ref.~\cite{BR}.
Further
the Relativistic Heavy Ion Collider (RHIC) has started
to measure several physical processes in hot matter
which include the dilepton energy spectra.
Therefore it is interesting to study the vector meson mass
including the {\it all possible thermal effects},
which is one of the important quantities
in the chiral phase transition.

For studying the vector meson mass in hot matter,
it is convenient to use a model
including the vector meson in a manner consistent with the 
chiral symmetry.
One of such models is the model based on 
the hidden local symmetry (HLS)
which successfully describes the systems including 
the pions and vector mesons at zero temperature~\cite{BKUYY,BKY:PR,HY:PR}.
In the framework of HLS, the vector meson is introduced
as the gauge boson into the system and acquires its mass
through the Higgs mechanism.
Based on the chiral perturbation theory (ChPT) with HLS,
the Wilsonian matching was proposed~\cite{HY:WM}. 
This is a manner which determines the parameters of 
the HLS Lagrangian from the underlying QCD 
at the matching scale $\Lambda$.
Recently in Ref.~\cite{HY:VM}, by using the Wilsonian matching,
the vector manifestation (VM) was proposed
as a novel manifestation of the chiral symmetry in the Wigner
realization, in which the chiral symmetry is restored by 
the massless degenerate pion (and its flavor partners) and
vector meson (and its flavor partners) as the chiral partner.
(For a review of the ChPT with HLS, the Wilsonian matching
and VM, see Ref.~\cite{HY:PR}.)

In Ref.~\cite{HS:PLB}, 
we extended the Wilsonian matching to the one at non-zero temperature
and showed that the VM actually occurs at the critical temperature
of the chiral symmetry restoration.
There, the {\it intrinsic temperature dependences} 
of the parameters of the HLS Lagrangian play the essential roles
to realize the chiral symmetry restoration consistently:
In the framework of the HLS the equality between 
the axial-vector and vector current correlators at critical point
can be satisfied only if the intrinsic thermal effects are
included.
Since the VM is a new picture which realizes the dropping mass of the
vector meson such as the one predicted by the 
Brown-Rho scaling~\cite{BR},
it is important to study what the VM predicts on the
properties of the pion and vector mesons in hot matter.

In this talk,
we shed a light on the validity of the vector dominance (VD)
in hot matter.
In several analyses such as the one on the dilepton spectra 
in hot matter done in Ref~\cite{Rapp-Wambach:00}, 
the VD is assumed to be held even in the high temperature region.
However, 
the analysis in Ref.~\cite{Pisarski}, which shows
that the thermal vector meson mass goes up if the VD holds,
seems to imply that the assumption of the VD
excludes the possibility
of the dropping mass of the vector meson from the beginning.
In the present analysis,
we present a new prediction of the VM in hot matter
on the direct photon-$\pi$-$\pi$ coupling which measures the validity
of the VD of the electromagnetic form factor of the pion. 
We find that {\it the VM predicts the large violation of the VD at
the critical temperature}~\cite{HS:VD}.
This indicates that the assumption of 
the VD may need to be weakened, at least in some amounts,
for consistently including the
effect of the dropping mass of the vector meson.

\section{Hidden Local Symmetry}

In this section, we briefly review 
the model based on the hidden local symmetry (HLS).

The HLS model is based on 
the $G_{\rm{global}} \times H_{\rm{local}}$ symmetry,
where $G=SU(N_f)_L \times SU(N_f)_R$ is the chiral symmetry
and $H=SU(N_f)_V$ is the HLS. 
The basic quantities are 
the HLS gauge boson $V_\mu$ and two matrix valued
variables $\xi_L(x)$ and $\xi_R(x)$
which transform as
\begin{equation}
  \xi_{L,R}(x) \to \xi^{\prime}_{L,R}(x)
  =h(x)\xi_{L,R}(x)g^{\dagger}_{L,R}\ ,
\end{equation}
where $h(x)\in H_{\rm{local}}\ \mbox{and}\ g_{L,R}\in
[\mbox{SU}(N_f)_{\rm L,R}]_{\rm{global}}$.
These variables are parameterized as
\begin{equation}
  \xi_{L,R}(x)=e^{i\sigma (x)/{F_\sigma}}e^{\mp i\pi (x)/{F_\pi}}\ ,
\end{equation}
where $\pi = \pi^a T_a$ denotes the pseudoscalar Nambu-Goldstone bosons
associated with the spontaneous symmetry breaking of
$G_{\rm{global}}$ chiral symmetry, 
and $\sigma = \sigma^a T_a$ denotes
the Nambu-Goldstone bosons associated with 
the spontaneous breaking of $H_{\rm{local}}$.
This $\sigma$ is absorbed into the HLS gauge 
boson through the Higgs mechanism,
and then the vector meson acquires its mass.
$F_\pi \ \mbox{and}\ F_\sigma$ are the decay constants
of the associated particles.
The phenomenologically important parameter $a$ is defined as 
\begin{equation}
  a = \frac{{F_\sigma}^2}{{F_\pi}^2}\ .
\end{equation}
The covariant derivatives of $\xi_{L,R}$ are given by
\begin{eqnarray}
 D_\mu \xi_L &=& \partial_\mu\xi_L - iV_\mu \xi_L + i\xi_L{\cal{L}}_\mu,
 \nonumber\\
 D_\mu \xi_R &=& \partial_\mu\xi_R - iV_\mu \xi_R + i\xi_R{\cal{R}}_\mu,
\end{eqnarray}
where $V_\mu$ is the gauge field of $H_{\rm{local}}$, and
${\cal{L}}_\mu \ \mbox{and}\ {\cal{R}}_\mu$ are the external
gauge fields introduced by gauging $G_{\rm{global}}$ symmetry.

The HLS Lagrangian with lowest derivative terms at the chiral limit
is given by~\cite{BKUYY,BKY:PR}
 \begin{equation}
  {\cal{L}} = {F_\pi}^2\mbox{tr}\bigl[ \hat{\alpha}_{\perp\mu}
                                      \hat{\alpha}_{\perp}^{\mu}
                                   \bigr] +
       {F_\sigma}^2\mbox{tr}\bigl[ \hat{\alpha}_{\parallel\mu}
                  \hat{\alpha}_{\parallel}^{\mu}
                  \bigr] -
        \frac{1}{2g^2}\mbox{tr}\bigl[ V_{\mu\nu}V^{\mu\nu}
                   \bigr]
\ , \label{eq:L(2)}
 \end{equation}
where $g$ is the HLS gauge coupling,
$V_{\mu\nu}$ is the field strength
of $V_\mu$ and
 \begin{eqnarray}
  \hat{\alpha}_{\perp }^{\mu}
     &=& \frac{1}{2i}\bigl[ D^\mu\xi_R \cdot \xi_R^{\dagger} -
                          D^\mu\xi_L \cdot \xi_L^{\dagger}
                   \bigr] \ ,
\nonumber\\
  \hat{\alpha}_{\parallel}^{\mu}
     &=& \frac{1}{2i}\bigl[ D^\mu\xi_R \cdot \xi_R^{\dagger}+
                          D^\mu\xi_L \cdot \xi_L^{\dagger}
                   \bigr]
\ .
 \end{eqnarray}
Expanding the Lagrangian~(\ref{eq:L(2)}) in terms of the $\pi$ field 
with taking the unitary gauge of the HLS ($\sigma = 0$),
we find the expressions for the mass of vector meson $M_\rho$,
$\rho\pi\pi$ coupling $g_{\rho\pi\pi}$,
$\rho$-$\gamma$ mixing strength $g_\rho$ and
direct $\gamma\pi\pi$ coupling $g_{\gamma\pi\pi}$.
Especially $g_{\gamma\pi\pi}$ coupling constant is expressed as
\begin{eqnarray}
 g_{\gamma\pi\pi} = e\,\Bigl( 1-\frac{a}{2} \Bigr).
\end{eqnarray}
By taking the parameter choice $a=2$,
the HLS reproduces the following 
phenomenological facts~\cite{BKUYY}:
$g_{\rho\pi\pi}=g$ (universality of the $\rho$ coupling)~\cite{Sakurai};
${M_\rho}^2=2g_{\rho\pi\pi}^2{F_\pi}^2$ (the KSRF relation, version II)
~\cite{KSRF};
$g_{\gamma\pi\pi}=0$ (vector dominance of the electromagnetic 
form factor of the pion)~\cite{Sakurai}.
Note that in framework of the HLS model, 
the vector dominance (VD) is satisfied
by the choice $a = 2$.

In the HLS model it is possible to perform the derivative
expansion systematically~\cite{Georgi,Tana,HY:PR}.
In this ChPT with  HLS the
vector meson mass is considered as small
compared with the chiral symmetry breaking scale 
$\Lambda_\chi$, by assigning ${\cal O}(p)$ to 
the HLS gauge coupling~\cite{Georgi,Tana}: 
\begin{equation}
 g \sim {\cal O}(p).
\end{equation}
For details of the ChPT with HLS, see Ref.~\cite{HY:PR}.

\section{Wilsonian Matching in Hot Matter}

When we naively extended a result obtained in the low
temperature region to the higher temperature region,
the axial-vector and vector current correlators 
do not agree with each other at the critical temperature.
Disagreement between these current correlators
is obviously inconsistent with 
the chiral symmetry restoration in QCD.
However the parameters of the HLS Lagrangian should be determined by
the underlying QCD.  
Thus it is natural that these parameters are dependent on temperature.
In Ref.~\cite{HS:PLB} the Wilsonian matching, which was originally
proposed at $T=0$~\cite{HY:WM}, was extended to non-zero temperature
and it was shown that the parameters of the HLS Lagrangian have the
intrinsic temperature dependences.
Further in Ref.~\cite{HKRS}, it was shown that the effects of Lorentz
symmetry violation at bare level are small
through the Wilsonian matching at non-zero temperature.
As is stressed in Ref.~\cite{HS:PLB}, the disagreement between
the axial-vector and vector current correlators 
mentioned above is cured by
including the intrinsic temperature dependences of the parameters of
the HLS Lagrangian.

The Wilsonian matching proposed in Ref.~\cite{HY:WM} is done by matching
the axial-vector and vector current correlators derived from the
HLS with those by the operator product expansion (OPE) in
QCD at the matching scale $\Lambda$.
In Refs.~\cite{HS:PLB,HKRS}, this matching scheme was extended to
the one at non-zero temperature.
Since the Lorentz symmetry breaking effect in
the bare pion decay constant is small, 
$F_{\pi,\rm{bare}}^t \simeq F_{\pi,\rm{bare}}^s$~\cite{HKRS},
it is a good approximation that we determine the bare pion decay
constant at non-zero temperature through the matching 
condition at zero temperature with putting possible 
temperature dependences on the gluonic 
and quark condensates~\cite{HS:PLB, HKRS}:
\begin{equation}
 \frac{F^2_\pi (\Lambda ;T)}{{\Lambda}^2} 
  = \frac{1}{8{\pi}^2}\Bigl[ 1 + \frac{\alpha _s}{\pi} +
     \frac{2{\pi}^2}{3}\frac{\langle \frac{\alpha _s}{\pi}
      G_{\mu \nu}G^{\mu \nu} \rangle_T }{{\Lambda}^4} +
     {\pi}^3 \frac{1408}{27}\frac{\alpha _s{\langle \bar{q}q
      \rangle }^2_T}{{\Lambda}^6} \Bigr]
\ .
\label{eq:WMC A}
\end{equation}
Through this condition
the temperature dependences of the quark and gluonic condensates
determine the intrinsic temperature dependences 
of the bare parameter $F_\pi(\Lambda;T)$,
which is then converted into 
those of the on-shell parameter $F_\pi(\mu=0;T)$ 
through the Wilsonian RGEs.

Now we consider the Wilsonian matching near the critical temperature
with assuming that the quark condensate becomes zero for $T \to T_c$.
As was discussed in Ref.~\cite{HKRS,HS:VD}, 
we can use the Lorentz invariant Lagrangian at bare level 
as long as we study the pion decay constant and validity of VD.
Then we start from the Lorentz invariant bare Lagrangian
even in hot matter.
The agreement between the current correlators,
which characterizes the chiral symmetry restoraion, 
$G_A^{(\rm{HLS})}=G_V^{(\rm{HLS})}$ is satisfied
only if the following conditions are met~\cite{HS:PLB}: 
\begin{eqnarray}
 g(\Lambda ;T) &\stackrel{T \to T_c}{\to}& 0, \nonumber\\
 a(\Lambda ;T) &\stackrel{T \to T_c}{\to}& 1. \label{eq:VM}
\end{eqnarray}
Note that $(g,a)=(0,1)$ is the fixed point of RGEs 
for $g$ and $a$~\cite{HY:WM}.
This implies that the parameter $M_\rho$ goes to zero for $T \to T_c$:
\begin{equation}
 M_\rho \stackrel{T \to T_c}{\to} 0.
\end{equation}

Including the hadronic thermal effects as well as 
quantum corrections through the RGEs, 
we obtain the physical quantities.
Near the critical temperature, we find that
the pole mass of the vector meson is the form
\begin{equation}
 m_\rho^2(T) \simeq M_\rho^2 \bigl[ 1 + \delta_{\rm{(had)}}(T) \bigr],
\end{equation}
where the hadronic correction is obtained as
\begin{equation}
 \delta_{\rm{(had)}}(T)
  = N_f \frac{15 - a^2}{144\,a} \frac{1}{F_\pi^2} T^2
  > 0.
\end{equation}
i.e., the hadronic thermal effect gives a positive correction near $T_c$.
However the pole mass $m_\rho$ goes to zero by vanishing parameter 
$M_\rho$ at the critical temperature and the vector manifestation (VM) 
of chiral symmetry is realized~\cite{HS:PLB}.

\section{Vector Manifestation of Chiral Symmetry}

The VM is highly in contrast with the standard chiral symmetry
restoration based on the linear sigma model.
In order to clarify this difference, 
we consider the multiplet structure.
In the broken phase, the chiral representation does not agree with
the mass eigenstate and there exists a mixing.
Then the scalar, pseudoscalar, vector and axial-vector mesons
belong to the following representations for $N_f=3$ respectively:
\begin{eqnarray}
 |s \rangle 
   &=& |(3,3^*) \oplus (3^*,3) \rangle, \nonumber\\
 |\pi \rangle 
   &=& |(3,3^*) \oplus (3^*,3) \rangle \sin\psi +
       |(1,8) \oplus (8,1) \rangle \cos\psi, \nonumber\\
 |\rho \rangle 
   &=& |(1,8) \oplus (8,1) \rangle, \nonumber\\
 |A_1 \rangle 
   &=& |(3,3^*) \oplus (3^*,3) \rangle \cos\psi -
       |(1,8) \oplus (8,1) \rangle \sin\psi,
\end{eqnarray}
where $\psi$ denotes the mixing angle 
and is given by $\psi \simeq 45\,{}^\circ$
~\cite{Gilman:1967qs,Weinberg:hw}.

Now we consider the chiral symmetry restoration, 
where it is expected that the above mixing disappears.
There are two possibilities of chiral symmetry realization.
One possible pattern is the case $\cos\psi \to 0$ for $T \to T_c$.
In this case, the pion belongs to $|(3,3^*) \oplus (3^*,3)\rangle$
and becomes the chiral partner of the scalar meson.
The vector and axial-vector mesons are in the same multiplet
$|(1,8) \oplus (8,1) \rangle$.
This is of course the standard scenario of chiral symmetry restoration.
Another possibility is the case $\sin\psi \to 0$ 
for $T \to T_c$~\cite{HY:VM}.
In this case, the pion belongs to purely $|(1,8) \oplus (8,1)\rangle$
and so its chiral partner is the vector meson:
\begin{eqnarray}
 |\pi \rangle 
   &=& |(1,8) \oplus (8,1) \rangle, \nonumber\\
 |\rho \rangle 
   &=& |(1,8) \oplus (8,1) \rangle
 \qquad \mbox{for}\quad \sin\psi \to 0.
\end{eqnarray}
The scalar meson joins with the axial-vector meson 
in the same representation $|(3,3^*) \oplus (3^*,3) \rangle$.
This is nothing but the VM of chiral symmetry.

\section{Violation of the Vector Dominance}

As we mentioned in introduction, in Ref.~\cite{HS:VD}, 
we studied the validity of the vector dominance (VD) which describes
the phenomena in low energy region very well.
At non-zero temperature there exists the hadronic thermal correction
to the parameters.
Thus it is nontrivial whether or not the
VD is realized in hot matter,
especially near the critical temperature.
Here we will show that the intrinsic temperature dependences
of the parameters of the HLS Lagrangian play essential roles, 
and then the VD is largely violated near the critical temperature.

We first mention the direct $\gamma\pi\pi$ interaction at zero
temperature. 
At the leading order of the derivative expansion in the HLS,
we can read
the form of the direct $\gamma\pi\pi$ interaction from
Eq.~(\ref{eq:L(2)}) as
\begin{equation}
 \Gamma_{\gamma\pi\pi{\rm(tree)}}^\mu 
  = e(q - k)^\mu(1-\frac{a}{2})\ ,
            \label{eq:gpp}
\end{equation}
where 
$e$ is the electromagnetic coupling constant and
$q$ and $k$ denote outgoing momenta of the pions. 
As we have mentioned in section 2,
for the parameter choice $a=2$
the direct $\gamma\pi\pi$ coupling vanishes, which leads to
the vector dominance of the electromagnetic form factor of the pion.

Next we evaluate the $\gamma\pi\pi$ coupling including quantum
corrections as well as hadronic thermal effects.
We obtained the parameters $a^t(T)$ and $a^s(T)$ which are the extension
of the parameter $a$ in hot matter,
in low temperature region $(T \ll M_\rho)$ as follows:
\begin{equation}
 a^t(T) \simeq a^s(T) 
        \simeq
          a(0) \left[
          1 + \frac{N_f}{12} \left( 1 - \frac{a^2}{4 a(0)} \right)
          \frac{T^2}{F_\pi^2(0;T)} \right],
  \label{at as form low T}
\end{equation}
where $a$ is the parameter renormalized at the scale $\mu=M_\rho$,
while $a(0)$ is the parameter defined from the $\gamma\pi\pi$
interaction at one-loop order. 
By using $F_\pi(0)=86.4\,\mbox{MeV}$,
$a(0) \simeq 2.31$ and
$a(M_\rho) = 1.38$ 
obtained through the Wilsonian matching for
$(\Lambda_{\rm QCD}\,,\,\Lambda) = (0.4\,,\,1.1)\,\mbox{GeV}$
and $N_f = 3$~\cite{HY:PR},
$a^t$ and $a^s$ in Eq.~(\ref{at as form low T}) are evaluated as
\begin{equation}
a^t(T) \simeq a^s(T) 
 \simeq
  a(0) \left[
  1 + 0.066 \left( \frac{T}{50\,\mbox{MeV}} \right)^2 
\right]
\ .
\end{equation}
This implies that the parameters $a^t$ and $a^s$ increase with
temperature in the low temperature region.
However, since the correction is small, we conclude that the
VD is still satisfied in the low temperature region.

When we consider the situation near the critical temperature,
the intrinsic thermal effects are important.
From the VM conditions in hot matter~(\ref{eq:VM}),
the parameters $(g,a)$ approach $(0,1)$
for $T \to T_c$ by the intrinsic temperature dependences.
Taking the limit $T \to T_c$, we find that
\begin{equation}
  a^t(T), \, a^s(T) 
   \stackrel{T \to T_c}{\to} 1 \ .
\end{equation}
This implies that the vector dominance is largely
violated near the critical temperature.

\section{Summary}

In the picture based on the vector manifestation (VM),
we presented a new prediction associated with
the validity of vector dominance (VD) in hot matter~\cite{HS:VD}.
In the HLS model at zero temperature, 
the Wilsonian matching predicts $a \simeq2$~\cite{HY:WM,HY:PR}
which guarantees the VD of the electromagnetic 
form factor of the pion.
Even at non-zero temperature,
this is valid as long as we consider the thermal effects
in the low temperature region,
where the intrinsic temperature dependences are negligible.
However the situation is changed when we consider the validity
of the VD in higher temperature region.
We showed that, {\it as a consequence of including the intrinsic effect,
the VD is largely violated at the critical temperature.}
In general, full temperature dependences include both hadronic and
intrinsic thermal effects.
Then there exist the violations of VD and universality of
the $\rho$-coupling at generic temperature,
although at low temperature the VD and universality are 
approximately satisfied.

In several analyses such as the one on the dilepton spectra 
in hot matter done in Ref~\cite{Rapp-Wambach:00}, 
the VD is assumed to be held even in
the high temperature region.
We should note that the analysis in Ref.~\cite{Pisarski} shows
that, if the VD holds, the thermal vector meson mass goes up.
Then the assumption of the VD, from the beginning, 
seems to exclude the possibility
of the dropping mass of the vector meson 
such as the one predicted by the 
Brown-Rho scaling~\cite{BR}.
Our result, which is consistent with the result in
Ref.~\cite{Pisarski} in some sense, indicates that the assumption of 
the VD may need to be weakened, at least in some amounts,
for consistently including the
effect of the dropping mass of the vector meson into the analysis.

\section*{Acknowledgment}

I would like to appreciate many discussions with
Professor Masayasu Harada, Doctor Youngman Kim, 
Professor Mannque Rho and Professor Koichi Yamawaki 
for our works.

\end{document}